# Signature of superconductivity in pressurized $La_4Ni_3O_{10}$

Qing Li, Ying-Jie Zhang, Zhe-Ning Xiang, Yuhang Zhang, Xiyu Zhu, Hai-Hu Wen[*]

National Laboratory of Solid State Microstructures and Department of Physics, Collaborative Innovation Center of Advanced Microstructures, Nanjing University, Nanjing 210093, China
*Corresponding authors: hhwen@nju.edu.cn

**Abstract:**

The discovery of high-temperature superconductivity near 80 K in bilayer nickelate $La_3Ni_2O_7$ under high pressures has renewed the exploration of superconducting nickelate in bulk materials. The extension of superconductivity in other nickelates in a broader family is also essential. Here, we report the experimental observation of superconducting signature in trilayer nickelate $La_4Ni_3O_{10}$ under high pressures. By using a modified sol-gel method and post-annealing treatment under high oxygen pressure, we successfully obtained polycrystalline $La_4Ni_3O_{10}$ samples with different transport behaviors at ambient pressure. Then we performed high-pressure electrical resistance measurements on these samples in a diamond-anvil-cell (DAC) apparatus. Surprisingly, the signature of possible superconducting transition with a maximum transition temperature ($T_c$) of about 20 K under high pressures is observed, as evidenced by a clear drop of resistance and the suppression of resistance drops under magnetic fields. Although the resistance drop is sample-dependent and relatively small, it appears in all of our measured samples. We argue that the observed superconducting signal is most likely to originate from the main phase of $La_4Ni_3O_{10}$. Our findings will motivate the exploration of superconductivity in a broader family of nickelates and shed light on the understanding of the underlying mechanisms of high-$T_c$ superconductivity in nickelates.

**Introduction**

The discovery of superconductivity in $La_{2-x}Ba_xCuO_4$ has opened the avenue to high-temperature superconductivity (HTS) in cuprates [1]. After the efforts of more than three decades, we know that the superconductivity in cuprates occurs in a wide variety of crystal structures, which include octahedral, square-planar, and pyramidal geometries of copper-oxygen networks [2-4]. In all cases, the two-dimensional $CuO_2$ plane and half-filled Cu $3d^9$ electronic configuration are believed to be essential to the emergence of high-$T_c$ superconductivity [5, 6]. Due to the striking structural and electronic similarities with cuprates, nickelates have been seen as an ideal candidate for exploring new high-$T_c$ superconductivity [7, 8]. In 2019, Li et al. reported the observation of superconductivity in $Nd_{1-x}Sr_xNiO_2$ thin films with $T_c$ around 9-15 K [9]. Shortly after that, superconductivity has been extended to other hole-doped infinite-layer nickelates with different rare-earth elements and $Nd_6Ni_5O_{12}$ with quintuple $NiO_2$ layers [10-17]. Under high pressures, the maximum $T_c$ of 31 K in $Pr_{0.8}Sr_{0.2}NiO_2$ film is observed [18]. However, in bulk infinite-layer nickelates, superconductivity is still absent [19, 20].

The structures of superconducting nickelate thin film all belong to the square-planar structure, which can be obtained from the pristine Ruddlesden-Popper (RP) phase by removing the apical oxygen [21, 22]. The RP phase with the chemical formula $R_{n+1}Ni_nO_{3n+1}$ ($R$ is Rare-earth elements; $n$ = 1, 2, 3 and ∞) are a huge family of materials and have been studied extensively [23-34]. Very recently, signatures of superconductivity up to nearly 80 K were reported in $La_3Ni_2O_7$ single crystals over 14 GPa [35]. The observation is in contrast to the infinite-layer nickelates since the existence of apical oxygens will lead to a nominal oxidation state of $Ni^{2.5+}(3d^{7.5})$, and the interlayer coupling between Ni-$3d_{z^2-r^2}$ and O-$p_z$ orbitals is much stronger than that of the infinite-layer superconductors. The crystal structure investigation on $La_3Ni_2O_7$ reveals a structure phase transition from the *Amam* space group to the *Fmmm* space group at around 15 GPa, and the high-pressure structure with the apical Ni-O-Ni bond approaching 180° is argued to be responsible for the superconductivity [35]. After this work, several experimental and theoretical works have been carried out to investigate the underlying mechanisms of high-$T_c$ superconductivity in $La_3Ni_2O_7$, [36-52] and zero resistance was realized in both single crystal and polycrystalline samples under improved hydrostatic pressure environment [36, 37, 40]. Among RP series compounds, the trilayer nickelates $R_4Ni_3O_{10}$ ($R$ = La, Pr, Nd) have attracted more attention since it has been reported to undergo metal-to-metal transitions [25-29, 53-56] and the transition was ascribed to intertwined charge and spin density waves [57].

High-$T_c$ superconductors, such as cuprates and iron-based superconductors, can always form a broader family of materials with different crystal structures and $T_c$ values [58, 59]. To investigate whether other RP phases are also superconductive, in this work, we synthesized high-quality polycrystalline samples of $La_4Ni_3O_{10}$ with a modified sol-gel method and post-annealing treatment. The transport measurements under high pressure on both pristine and high oxygen pressure annealed samples reveal a superconducting-like resistance drop at around 15-20 K. And the resistance drop can be gradually suppressed by applying external magnetic fields, which further confirms the existence of superconductivity. After collecting systematic experiment results, we depict a temperature-pressure phase diagram of $La_4Ni_3O_{10}$, which reveals a dome-like superconducting region. The different critical pressures required for the appearance of superconductivity in different samples and relatively small resistance drop ratio (about 7%) both indicate that the sample quality or hydrostatic pressure conditions need to be further improved in future work.

**Experimental Methods**

The polycrystalline samples of $La_4Ni_3O_{10}$ were synthesized via a modified sol-gel method [60] and post-sintering treatment. A stoichiometric amount of high-purity $La_2O_3$ (Alfa, 99.99%) and NiO (Alfa, 99.99%) were dissolved in nitric acid. Then the mixture was heated to 150 °C, and an equivalent molar proportion of citric acid concerning cations in the mixture was added when $La_2O_3$ and NiO were dissolved. After several hours of heating and drying off, a vivid green gel was formed. The green product was transformed into gray powder after heating at 300°C for 5 hours. Then, the gray powder was ground and heated to 1080 °C and maintained at this temperature for 5 days in the air. After slowly cooling down with the furnace, we obtained the as-grown sample (S1) with a black color. Then we conducted the post-sintering treatment of S1 under high oxygen pressure using a piston-cylinder-type high-pressure apparatus (LP 1000-540/50, Max Voggenreiter) [19]. During the high-pressure processing, the powder of AgO with mass = 0.1 g was used for the purpose of oxidation. A tablet made of S1 with a weight of 0.1 g was separated from the AgO tablet by a BN pellet and sealed into a gold capsule. Then it was heated to 500 °C and held for 5 h at 2 GPa. Finally, we got a sample (S2) with better conductivity. For S3, we changed the raw material from NiO to $Ni(OH)_2$, since the latter dissolves more easily in nitric acid, and the other synthesis process is the same as S1.

The crystal structure of $La_4Ni_3O_{10}$ was identified by powder X-ray diffraction (XRD) with Cu Kα radiation (Bruker; *D*8 Advance diffractometer; λ =1.541 Å), the Rietveld refinements were conducted with TOPAS 4.2 software [61]. Temperature-dependent resistance measurements were carried out with a physical-property measurement system (PPMS-9 T, Quantum design). Diamond anvil cells (DACPPMS-ET225, Shanghai Anvilsource Material Technology Co., Ltd) with culets of 300 μm and 200 μm were used to generate pressures up to 50 GPa and 75 GPa, respectively. Four-probe van der Pauw method was adopted for the high-pressure resistance measurements. The ruby fluorescence method was used to detect the pressure at room temperature [62].

**Results and discussion**

The crystal structure of $La_4Ni_3O_{10}$ is shown in Figure 1(a), which belongs to the RP series compounds with *n* = 3 [21, 26, 29, 56]. The structure can be described as the stacking of perovskite $(LaNiO_3)_3$ blocks and rock salt (La-O) layers along the *c*-axis. Note that the ambient crystal structure of $La_4Ni_3O_{10}$ is quite similar to that of $La_3Ni_2O_7$, but the nominal valence state of $Ni^{2.67+}$ is a little larger than that of $La_3Ni_2O_7$ ($Ni^{2.5+}$). The crystal structure of $La_4Ni_3O_{10}$ at ambient pressure is still under debate and there are several different space groups proposed before, for example, *Imm2* [56], *Cmce (Bmab)* [63], $P2_1/a$ (*Z* = 2) [27], and $P2_1/a$ (*Z* = 4) [28, 29, 38]. Recently, Zhang *et al.* determined the crystal structure of $R_4Ni_3O_{10}$ (*R* = La, Pr) by using synchrotron and lab x-ray single crystal diffraction and found that the monoclinic structure $P2_1/a$ ($Z = 2$) is the room-temperature thermodynamic state [26].

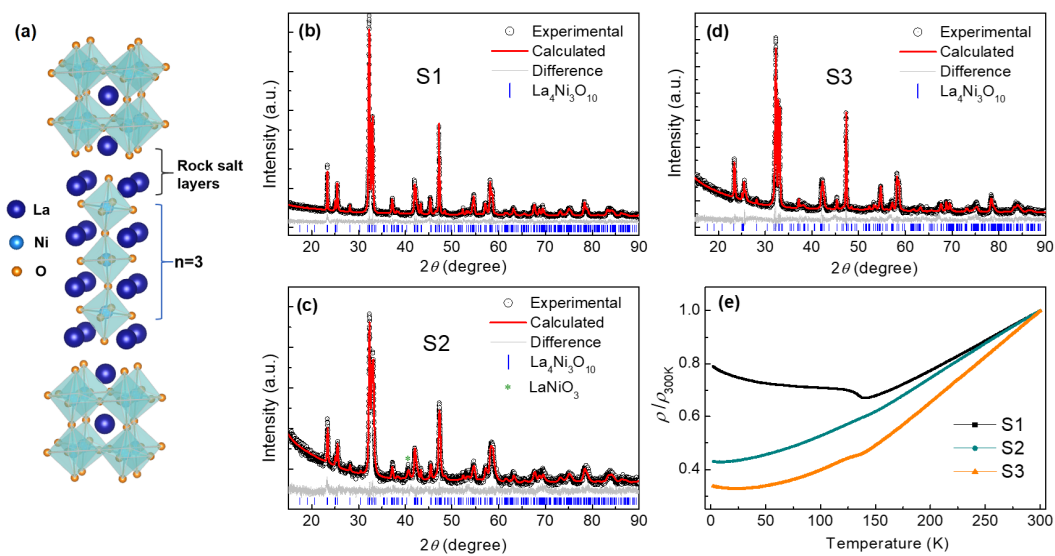

Fig. 1 (a) Schematic crystal structure of $La_4Ni_3O_{10}$. Lanthanum, nickel, and oxygen atoms are depicted in dark blue, cyan, and orange, respectively. (b-d) Powder x-ray diffraction patterns and its Rietveld fitting curves of $La_4Ni_3O_{10}$ (S1-S3). (e) Normalized temperature-dependent resistivity curves of $La_4Ni_3O_{10}$ (S1-S3) from 2 K to 300 K.

Figure 1(b-d) shows the powder XRD data and related Rietveld refinements of the $La_4Ni_3O_{10}$ samples with the space group of $P2_1/a$ ($Z = 2$). The purity of the $La_4Ni_3O_{10}$ phases is confirmed by indexing all the observed diffraction peaks from XRD data. For the post-annealed sample (S2), one tiny impurity peak at around 41° may arise from the $LaNiO_3$ phase. Details of structural parameters and Rietveld fitting data of all three samples are summarized in Table I. The lattice parameters are in good agreement with the previous reports [26, 53, 56]. The values of the agreement factor ($R_{wp}$ and $R_p$) are quite small, suggesting the high reliability of our Rietveld refinements.

**Table I** Crystallographic data obtained from the Rietveld refinements of $La_4Ni_3O_{10}$ (S1-S3) with the space group of $P2_1/a$ ($Z = 2$).

| Sample | S.G. | a (Å) | b (Å) | c (Å) | β (°) | Cell volume | $R_{wp}$(%) | $R_p$(%) | GOF |
|---|---|---|---|---|---|---|---|---|---|
| S1 | $P2_1/a$ | 5.4132(9) | 5.4627(6) | 14.2337(7) | 100.71(8) | 413.57(1) | 5.47 | 4.30 | 1.19 |
| S2 | $P2_1/a$ | 5.4131(6) | 5.4638(1) | 14.2461(6) | 101.26(1) | 413.23(9) | 4.91 | 3.77 | 1.14 |
| S3 | $P2_1/a$ | 5.4082(4) | 5.4533(6) | 14.2548(3) | 100.94(3) | 412.77(4) | 5.75 | 4.35 | 1.30 |

Figure 1(e) shows the normalized temperature-dependent resistivity of three La$_4$Ni$_3$O$_{10}$ samples (S1-S3). The resistivity decreases with decreasing temperature from 300 K, indicating a metallic behavior. And a density wave-like anomaly ($T^*$) was observed at about 134.9, 130.4, and 131.4 K for S1, S2, and S3, respectively. After this transition, as the temperature decreases further, the transport behaviors of the three samples are different from each other. For S1, the electrical resistance shows weak insulating behavior below the $T^*$, while S2 and S3 keep the metallic behavior after the transition. More specifically, from S1 to S3, the residual resistivity ratio of the material gradually increases. If we look back at the crystallographic data of La$_4$Ni$_3$O$_{10}$ (Table I), some clues may be found that the value of lattice parameter $c$ of the samples increases significantly from S1 to S3. From the previous literature [64, 65], we know that the oxygen stoichiometry has a great effect on the electronic properties of $R_4$Ni$_3$O$_{10}$ ($R$ = La, Pr) compounds. In this sense, there may be some underlying correlations between the lattice parameter $c$ and the oxygen stoichiometry in La$_4$Ni$_3$O$_{10}$, which needs to be clarified in future work.

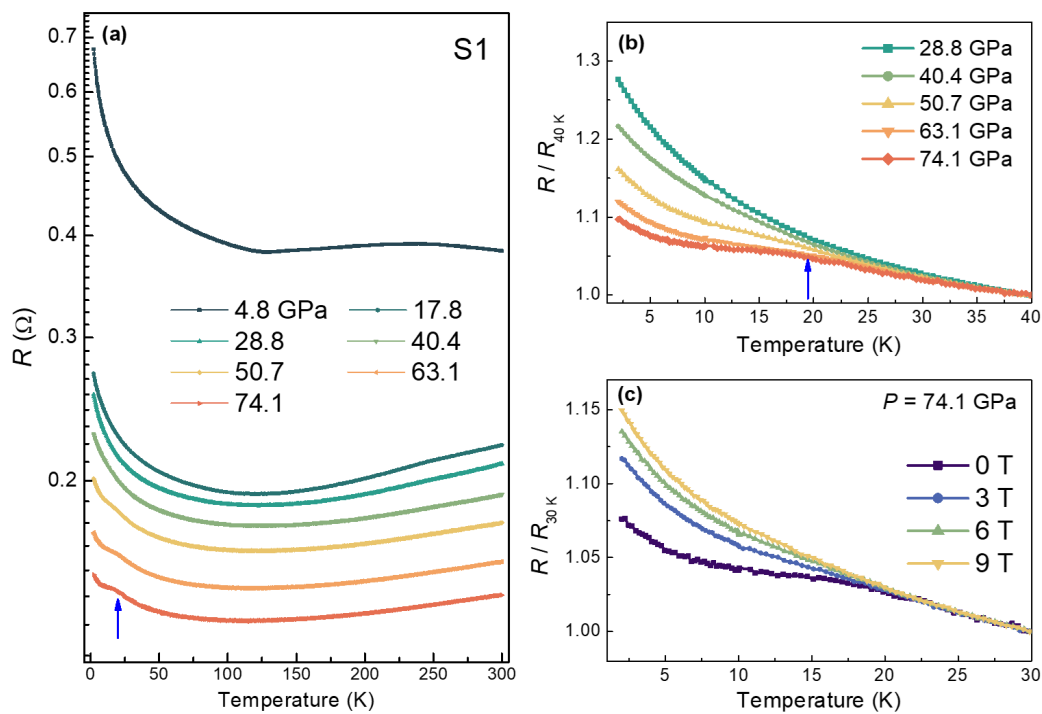

Fig. 2 (a) Temperature-dependent resistance ($R$ - $T$) curves of La$_4$Ni$_3$O$_{10}$ (S1) under various pressures up to 74.1 GPa. (b) Normalized $R/R_{40K}$ - $T$ curves from 2 to 40 K under selected pressures. (c) Temperature-dependent resistance measured at 74.1 GPa under various magnetic fields.

To figure out whether there is a superconducting transition under high pressures in $La_4Ni_3O_{10}$, we performed high-pressure electrical resistance measurements on S1 up to 74.1 GPa. As we can see from Fig. 2(a), the resistance values at room temperature decrease continuously with the increase of pressure, which can be attributed to the modification of the polycrystalline grain boundaries under compression. At low pressures, the temperature-dependent resistance (*R-T*) curves show similar behavior to that at ambient pressure, that is, the metallic behavior in the high-temperature region and the weak insulating upturn below $T^*$. For $P \geqslant 17$ GPa, the density wave-like transition ($T^*$) becomes broad and difficult to define from the *R-T* curves. However, to our surprise, a resistance anomaly at about 20 K appears at pressures above 50.7 GPa. Such a reduction of resistance becomes more and more pronounced with further increasing pressure as shown by the blue arrow in Fig. 2(a).

To show the low-temperature resistance anomaly more clearly, in Fig. 2 (b), we present the *R-T* curves normalized at 40 K under several pressures from 28.8 to 74.1 GPa. We can see that the low-temperature $R/R_{40K}$ begins deviating from the original weak insulating tendency when the pressure is higher than 50.7 GPa. Since the reduction in resistance usually corresponds to a superconducting transition in a material, we may attribute the observed resistance anomaly to pressure-induced superconductivity in our $La_4Ni_3O_{10}$ samples. Figure 2(c) presents the $R - T$ curves at 74.1 GPa under various magnetic fields. The gradual suppression of superconducting-like resistance anomaly under magnetic fields further confirmed the possible existence of superconductivity in our sample. A recent work reported by Zhang *et al.* [38] shows that superconductivity is not observed in as-grown polycrystalline $La_4Ni_3O_{10}$ up to about 50 GPa. The reason may be that the applied pressure has not reached the critical value for the occurrence of superconductivity or the difference in the sample itself.

From the literature [35-40], the observation of superconductivity in $La_3Ni_2O_7$ highly depends on the ground state of the sample at ambient pressure and zero resistance can only be achieved in a hydrostatic pressure condition with liquid pressure-transmitting-medium. Thus, we performed post-treatment on our as-grown sample (S1) under high oxygen pressures and obtained a sample (S2) with metallic behavior through the whole temperature range we measured as shown in Fig. 1(e). Figure 3 (a) shows the *R-T* curves of S2 under various pressures up to 47.5 GPa. To create a better hydrostatic environment, we employed soft KBr as the pressure-transmitting-medium in this experimental run. At low-pressure regions, the *R-T* curve of S2 exhibits a good metallic behavior and a weak resistance upturn below 10 K. The low-temperature resistance upturn in pressurized samples may be attributed to the grain boundary effect. As the pressure gradually increases, the low-temperature resistance upturn is weakened and a clear resistance drop is observed at around 34.5 GPa as shown by the blue arrow in Fig. 3(a). With further increasing pressure, the drop in resistance becomes more and more obvious, and the transport behavior above $T_c$ is metallic and shows a Fermi liquid behavior.

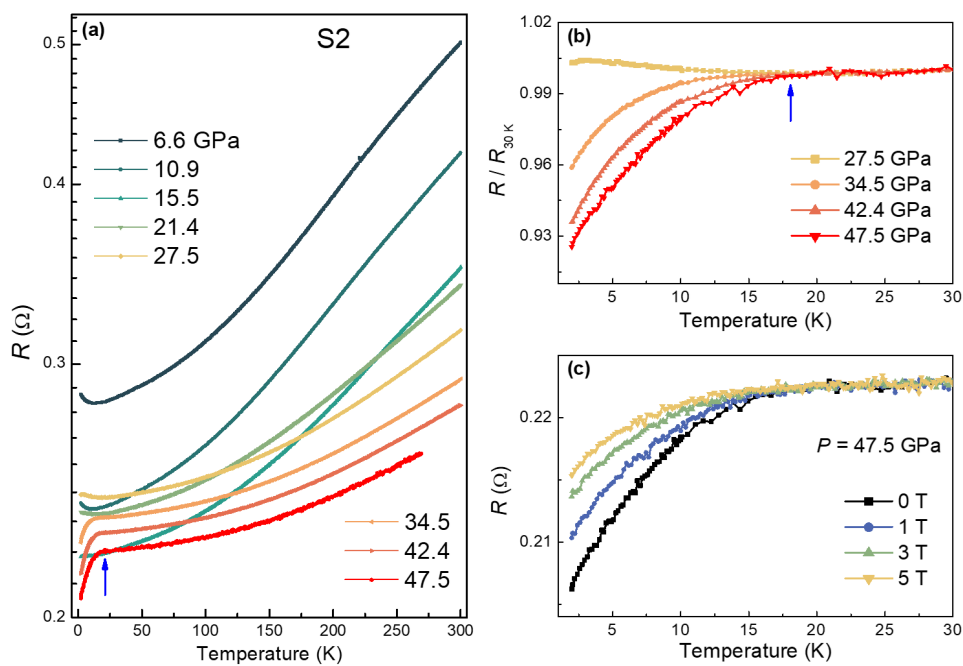

Fig. 3 (a) Temperature-dependent resistance (*R - T*) curves of S2 under various pressures up to 47.5 GPa. (b) Normalized $R/R_{30K}$ - *T* curves from 2 to 30 K at selected pressures. (c) Temperature-dependent resistance measured at 47.5 GPa under various magnetic fields.

Figure 3(b) shows the normalized $R/R_{30K}$ - *T* curves of S2 at pressures from 27.5 to 47.5 GPa. As we can see, from 30 to 20 K, the $R/R_{30K}$ - *T* curves almost coincide with each other under different pressures. Then, a clear drop in resistance appears below about 20 K, indicating a possible superconducting transition. The maximum value of the resistance drop ratio in our present run is about 7% under the pressure of 47.5 GPa. Figure 3(c) displays the $R$-$T$ curves of S2 in different applied magnetic fields at 47.5 GPa. The superconducting-like transition is gradually suppressed with increasing magnetic field, and the resistance drops survive even at a magnetic field of 5 T. From the experimental results above, we find that the post-annealing under high oxygen pressure can indeed optimize the transport behavior of the as-grown sample and get a better superconducting transition at low temperatures. However, at the current stage, we are still unable to get a complete superconducting transition, *i.e.,* zero resistance, even though the normal state behavior of $La_4Ni_3O_{10}$ is already a good metal. We believe that further optimization of the samples and the application of better hydrostatic pressures are essential. Very recently, during the preparation of the present work, we noticed a work by Sakakibara *et al.* [39] who claimed the occurrence of superconductivity in polycrystalline $La_4Ni_3O_{9.99}$. Their results are also based on the observation of a limited drop in the resistance at low temperatures but on a weak insulating background.

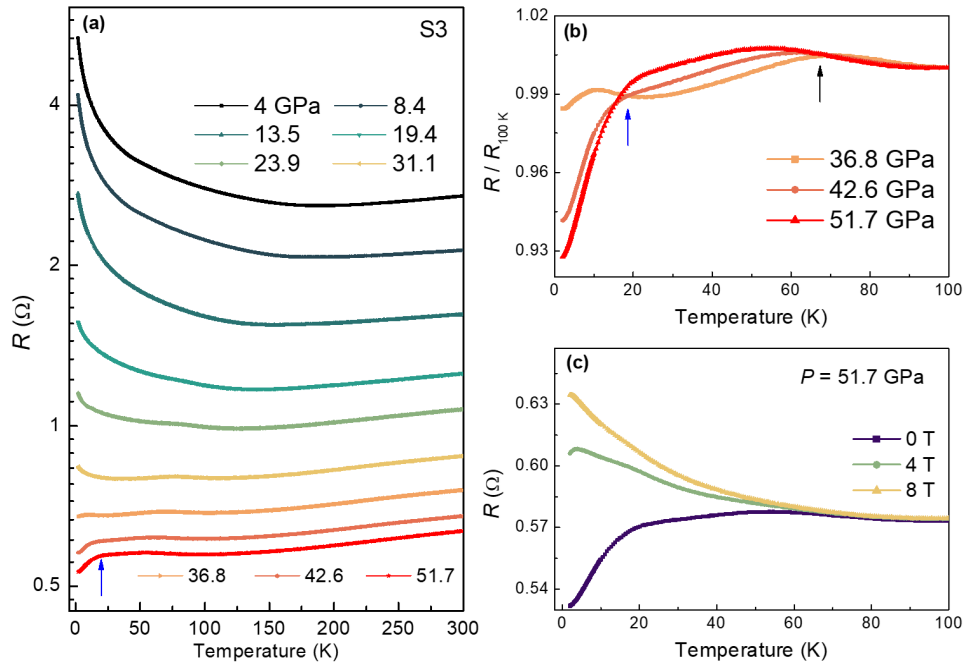

Fig. 4 (a) Temperature-dependent resistance ($R$ - $T$) curves of S3 under various pressure up to 51.7 GPa. (b) Normalized $R/R_{100K}$ - $T$ curves from 2 to 100 K at selected pressures. Two separated transitions as indicated by black and blue arrows are observed. (c) Temperature-dependent resistance measured at 51.7 GPa under various magnetic fields.

To investigate whether superconductivity in $La_4Ni_3O_{10}$ exists in different samples.

We then prepared a new $La_4Ni_3O_{10}$ (S3) using a modified sol-gel method with $Ni(OH)_2$ instead of NiO as raw material. As we can see from Fig. 1(e), S3 exhibits better metallic behavior than the first two samples (S1 and S2) at ambient pressure, and the metal-to-metal transition at around 140 K is also clearly observed. We then performed high-pressure resistance measurements on S3, and the results are presented in Fig. 4 (a). At low pressure (4 GPa), the ground state of S3 changes from metallic to a weakly insulating state at low temperatures, similar to the observation on S1 and previous reports on $La_3Ni_2O_7$ and other oxides [35, 36, 66]. With further increasing pressure, the low-temperature insulating behavior is gradually suppressed and a metallic behavior is realized at about 36.8 GPa. Once again, a resistance drop is also observed in low-temperature regions as shown by the blue arrow in Fig. 4(a). To our surprise, an additional resistance anomaly is observed in the high-temperature region, and this anomaly can be traced to lower pressures at about 19.4 GPa. The resistance drop or anomaly can be visualized more clearly in $R/R_{100K}$ - $T$ curves as shown in Fig. 4 (b). There are two transitions under high pressures above 36.8 GPa in S3. The low-temperature transition ($T_{c1}$) similar to the S1 and S2 occurs in the temperature region from 10 to 20 K and is gradually enhanced with increasing pressure. However, the high-temperature transition ($T_{c2}$) occurs at 60-80 K and decreases with increasing pressure. Furthermore, both transitions can be gradually suppressed by magnetic fields as shown in Fig. 4(c), indicating two possible superconducting transitions at low temperatures in S3. According to the previous literature [38, 39], we may attribute the $T_{c2}$ to the residual $La_3Ni_2O_7$ phase in our sample since the transition temperature and its pressure-dependent behaviors are quite similar.

Based on the above results, we can establish a $T$-$P$ phase diagram for $La_4Ni_3O_{10}$ as shown in Fig. 5. The metal-to-metal transitions ($T^*$) temperature is defined from the kink feature in $\rho$-$T$ curve around 140 K. The temperature at which the resistance deviates from the initial linear tendency is considered as the onset of superconducting transition ($T_c$). At ambient or low-pressure regions, a density wave-like transition can be observed and it seems to decrease with increasing pressure. At high pressures, a superconducting-like transition is observed at around 15-20 K, giving a dome-like superconducting region as shown by the solid circles with different colors in Fig. 5. We can also observe a high-temperature superconducting-like transition ($T_{c2}$) at around 70 K in S3 as shown by the open circles in Fig. 5, and the $T_c$ values and their pressure-dependent behaviors are quite similar to the previous observation in $La_3Ni_2O_7$. Thus, we may attribute it to the residual $La_3Ni_2O_7$ phase in S3 at the current stage, due to the similar synthesis conditions of these two phases [56].

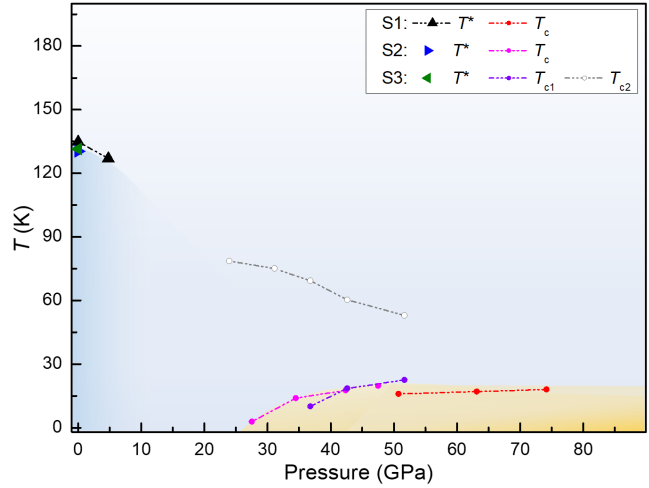

Fig. 5 Temperature-pressure phase diagram of $La_4Ni_3O_{10}$. The solid triangles with different colors represent the density wave-like transitions ($T^*$) at ambient and low pressures. The solid circles with different colors represent the onset superconducting transition temperatures ($T_c$) of three measured samples. The open circles represent the observed high-temperature transition ($T_{c2}$) in S3.

Because of the observation of superconductivity in bilayer nickelate $La_3Ni_2O_7$, the occurrence of superconductivity in $La_4Ni_3O_{10}$ seems to be expectable. According to the theoretical proposal from density functional theory calculations in ref. [39], trilayer nickelate $La_4Ni_3O_{10}$ may also become superconductive under high pressure with $T_c$ comparable to some cuprates but lower than $La_3Ni_2O_7$, and the $d_{3z^2-r^2}$ band also contributes to superconductivity around the stoichiometric composition in $La_4Ni_3O_{10}$. We also want to point out that the observed signature of superconductivity in our samples, *i.e.,* the drops of resistance and gradual suppression with increasing magnetic field, cannot give conclusive evidence for a superconductor. For the confirmation of bulk superconductivity, zero resistance and diamagnetism are both essential. Nonetheless, although the critical pressure for the emergence of superconductivity and the ratio of resistance drops are sample-dependent, the signatures appear in all of our measured samples and seem quite robust. In this sense, we may argue that the observed superconducting signal is most likely originated from the main phase of $La_4Ni_3O_{10}$.

As to why the superconductivity transition in $La_4Ni_3O_{10}$ under high pressures is so weak, we may attribute it to the following two possible reasons. First, the nonstoichiometric composition of oxygen in the sample. The variation of oxygen content in RP phase nickelates will have a significant effect on the physical properties of the material [35, 64, 65]. As we know from ref. [35], the slight deficiency of oxygen in $La_3Ni_2O_{7-\delta}$ will have a significant impact on the appearance of superconductivity at high pressures. We speculate that a similar situation may also be present in $La_4Ni_3O_{10}$. Second, hydrostatic pressure conditions are not realized in our high-pressure measurements. From theoretical predictions on $La_4Ni_3O_{10}$ [39], a structural phase transition from monoclinic to tetragonal is expected under high pressures. If the appearance of superconductivity in $La_4Ni_3O_{10}$ is closely associated with flatter $NiO_2$ planes (tetragonal symmetry), then the hydrostatic pressure conditions during the measurements are undoubtedly important. The same scenario has been confirmed in $La_3Ni_2O_7$. By using liquid pressure-transmitting-medium to produce better hydrostatic condition, high-$T_c$ superconductivity with clear zero resistance is observed in both single crystal and polycrystalline samples. [36, 37, 40] Unfortunately, the extreme hydrostatic pressure over 30 GPa is difficult to be achieved in labs. Based on the above discussion, if the high-pressure phase with flat $NiO_2$ planes is crucial for the occurrence of superconductivity, one feasible way is to reduce the critical pressure for the phase transition to occur. Chemical doping or the strain effects from thin films and heterostructures would be effective tools [9, 55]. If all go well, a superconducting RP phase with flat $NiO_2$ planes may be obtained at lower or even ambient pressure. Related experiments are highly desired and are actually underway.

**Conclusion**

In summary, we have successfully synthesized the polycrystalline samples of $La_4Ni_3O_{10}$ with a modified sol-gel method and post-annealing treatment. The crystal structure and transport measurements at ambient pressure confirmed the high quality of the obtained samples. Surprisingly, under high pressures, a possible superconducting transition occurs at about 20 K, as evidenced by the reduction or drop of resistance on the *R-T* curves and the magnetic field enhanced resistance in the low temperature region. The transition temperature shows also a dome-shaped region in $La_4Ni_3O_{10}$. The drops of resistance observed in our samples are relatively small in the current stage, but they appear in all of our measured samples and could be enhanced by further optimizing sample quality and hydrostatic pressure conditions in the future. Thus, we believe that the superconducting signal observed at high pressure in our samples can be attributed to the intrinsic property of $La_4Ni_3O_{10}$ itself. Our experimental observation can motivate the exploration of the superconductivity phenomenon with higher $T_c$ in the broad family of nickelates.

**Acknowledgments**

This work is supported by the National Key R&D Program of China (No. 2022YFA1403201), National Natural Science Foundation of China (Nos. 12204231, 12061131001, 52072170, and 11927809), and Strategic Priority Research Program (B) of Chinese Academy of Sciences (No. XDB25000000).

**Reference**

[1] J. G. Bednorz, and K. A. Müller, *Possible high $T_c$ superconductivity in the Ba–La–Cu–O system.* Z. Phys. B-Condens. Matter **64,** 189-193 (1986).

[2] H. Sawa, S. Suzuki, M. Watanabe, J. Akimitsu, H. Matsubara, H. Watabe, S. Uchida, K. Kokusho, H. Asano, F. Izumi, and E. Takayama-Muromachi, *Unusually simple crystal structure of an Nd-Ce-Sr-Cu-O superconductor.* Nature **337,** 347-348 (1989).

[3] M. G. Smith, A. Manthiram, J. Zhou, J. B. Goodenough, and J. T. Markert, *Electron-doped superconductivity at 40 K in the infinite-layer compound $Sr_{1-y}Nd_yCuO_2$.* Nature **351,** 549-551 (1991).

[4] K. Yvon, and M. Frangois, *Crystal structures of high-$T_c$ oxides.* Z. Phys. B-Condens. Matter **76,** 413-444 (1989).

[5] B. Keimer, S. A. Kivelson, M. R. Norman, S. Uchida, and J. Zaanen, *From quantum matter to high-temperature superconductivity in copper oxides.* Nature **518,** 179-186 (2015).

[6] P. A. Lee, N. Nagaosa, and X.-G. Wen, *Doping a Mott insulator: Physics of high temperature superconductivity.* Rev. Mod. Phys. **78,** 17-85 (2006).

[7] V. I. Anisimov, D. Bukhvalov, and T. M. Rice, *Electronic structure of possible nickelate analogs to the cuprates.* Phys. Rev. B **59,** 7901-7906 (1999).

[8] J. Chaloupka, and G. Khaliullin, *Orbital order and possible superconductivity in*

*LaNiO$_3$/LaMO$_3$ superlattices.* Phys. Rev. Lett. **100,** 016404 (2008).

[9] D. Li, K. Lee, B. Y. Wang, M. Osada, S. Crossley, H. R. Lee, Y. Cui, Y. Hikita, and H. Y. Hwang, *Superconductivity in an infinite-layer nickelate.* Nature **572,** 624-627 (2019).

[10] Q. Gu, Y. Li, S. Wan, H. Li, W. Guo, H. Yang, Q. Li, X. Zhu, X. Pan, Y. Nie, and H.-H. Wen, *Single particle tunneling spectrum of superconducting Nd$_{1-x}$Sr$_x$NiO$_2$ thin films.* Nat. Commun. **11,** 6027 (2020).

[11] M. Osada, B. Y. Wang, B. H. Goodge, K. Lee, H. Yoon, K. Sakuma, D. Li, M. Miura, L. F. Kourkoutis, and H. Y. Hwang, *A Superconducting Praseodymium Nickelate with Infinite Layer Structure.* Nano Lett. **20,** 5735-5740 (2020).

[12] S. W. Zeng, C. J. Li, L. E. Chow, Y. Cao, Z. T. Zhang, C. S. Tang, X. M. Yin, Z. S. Lim, J. X. Hu, P. Yang, and A. Ariando, *Superconductivity in infinite-layer nickelate La$_{1-x}$Ca$_x$NiO$_2$ thin films.* Sci. Adv. **8,** eabl9927 (2022).

[13] M. Osada, B. Y. Wang, B. H. Goodge, S. P. Harvey, K. Lee, D. Li, L. F. Kourkoutis, and H. Y. Hwang, *Nickelate Superconductivity without Rare-Earth Magnetism: (La,Sr)NiO$_2$.* Adv. Mater. **33,** 2104083 (2021).

[14] G. A. Pan, D. Ferenc Segedin, H. LaBollita, Q. Song, E. M. Nica, B. H. Goodge, A. T. Pierce, S. Doyle, S. Novakov, D. Cordova Carrizales, A. T. N`Diaye, P. Shafer, H. Paik, J. T. Heron, J. A. Mason, A. Yacoby, L. F. Kourkoutis, O. Erten, C. M. Brooks, A. S. Botana, J. A. Mundy, *Superconductivity in a quintuple-layer square-planar nickelate.* Nat. Mater. **21,** 160-164 (2022).

[15] W. Wei, D. Vu, Z. Zhang, F. J. Walker, C. H. Ahn, *Superconducting Nd$_{1-x}$Eu$_x$NiO$_2$ thin films using in situ synthesis.* Sci. Adv. **9,** eadh3327 (2023).

[16] X. Ding, C. C. Tam, X. Sui, Y. Zhao, M. Xu, J. Choi, H. Leng, J. Zhang, M. Wu, H. Xiao, X. Zu, M. Garcia-Fernandez, S. Agrestini, X. Wu, Q. Wang, P. Gao, S. Li, B. Huang, K.-J. Zhou, and L. Qiao, *Critical role of hydrogen for superconductivity in nickelates.* Nature **615,** 50-55 (2023).

[17] Q. Gu, and H.-H. Wen, *Superconductivity in nickel based 112 systems.* The Innovation **3,** 100202 (2022).

[18] N. N. Wang, M. W. Yang, Z. Yang, K. Y. Chen, H. Zhang, Q. H. Zhang, Z. H. Zhu, Y. Uwatoko, L. Gu, X. L. Dong, J. P. Sun, K. J. Jin, and J.-G. Cheng, *Pressure-induced monotonic enhancement of T$_c$ to over 30 K in superconducting Pr$_{0.82}$Sr$_{0.18}$NiO$_2$ thin films.* Nat. Commun. **13,** 4367 (2022).

[19] Q. Li, C. He, J. Si, X. Zhu, Y. Zhang, and H. H. Wen, *Absence of superconductivity in bulk Nd$_{1-x}$Sr$_x$NiO$_2$.* Commun. Mater. **1,** 16 (2020).

[20] B.-X. Wang, H. Zheng, E. Krivyakina, O. Chmaissem, P. P. Lopes, J. W. Lynn, L. C. Gallington, Y. Ren, S. Rosenkranz, J. F. Mitchell and D. Phelan, *Synthesis and characterization of bulk Nd$_{1-x}$Sr$_x$NiO$_2$ and Nd$_{1-x}$Sr$_x$NiO$_2$.* Phys. Rev. Mater. **4,** 084409 (2020).

[21] B. V. Beznosikov, and K. S. Aleksandrov, *Perovskite-Like Crystals of the Ruddlesden-Popper Series.* Crystallogr. Rep. **45,** 792 (2000).

[22] P. Lacorre, *Passage from T-Type to T`-Type Arrangement by Reducing R$_4$Ni$_3$O$_{10}$ to R$_4$Ni$_3$O$_8$ (R = La, Pr, Nd).* J. Solid State Chem. **97,** 495-500 (1992).

[23] S. Catalano, M. Gibert, J. Fowlie, J. Íñiguez, J.-M. Triscone, and J. Kreisel, *Rare-*

[23] *earth nickelates RNiO3: thin films and heterostructures.* Rep. Prog. Phys. **81,** 046501 (2018).

[24] J. Zhang, and X. Tao, *Review on quasi-2D square planar nickelates.* CrystEngComm, **23,** 3249-3264 (2021).

[25] G. Wu, J. J. Neumeier, and M. F. Hundley, *Magnetic susceptibility, heat capacity, and pressure dependence of the electrical resistivity of $La_3Ni_2O_7$ and $La_4Ni_3O_{10}$.* Phys. Rev. B **63,** 245120 (2001).

[26] J. Zhang, H. Zheng, Y.-S. Chen, Y. Ren, M. Yonemura, A. Huq, and J. F. Mitchell, *High oxygen pressure floating zone growth and crystal structure of the metallic nickelates $R_4Ni_3O_{10}$ (R = La, Pr).* Phys. Rev. Mater. **4,** 083402 (2020).

[27] S. Huangfu, G. D. Jakub, X. Zhang, O. Blacque, P. Puphal, E. Pomjakushina, F. O. von Rohr, and A. Schilling, Anisotropic character of the metal-to-metal transition in $Pr_4Ni_3O_{10}$. Phys. Rev. B **101,** 104104 (2020).

[28] B. Z. Li, C. Wang, P. T. Yang, J. P. Sun, Y. B. Liu, J. Wu, Z. Ren, J. G. Cheng, G. M. Zhang, and G. H. Cao, *Metal-to-metal transition and heavy-electron state in $Nd_4Ni_3O_{10-\delta}$.* Phys. Rev. B **101,** 195142 (2020).

[29] Q. Li, C. He, X. Zhu, J. Si, X. Fan, and H.-H. Wen, Contrasting physical properties of the trilayer nickelates $Nd_4Ni_3O_{10}$ and $Nd_4Ni_3O_8$. Sci. Chin. Phys. Mech. Astron. **64,** 227411 (2021).

[30] M. Huo, Z. Liu, H. Sun, L. Li, H. Lui, C. Huang, F. Liang, B. Shen, and M. Wang, *Synthesis and properties of $La_{1-x}Sr_xNiO_3$ and $La_{1-x}Sr_xNiO_2$.* Chin. Phys. B **31,** 107401 (2022).

[31] C. He, X. Ming, Q. Li, X. Zhu, J. Si and H.-H. Wen, *Synthesis and physical properties of perovskite $Sm_{1-x}Sr_xNiO_3$ (x = 0, 0.2) and infinite-layer $Sm_{0.8}Sr_{0.2}NiO_2$ nickelates.* J. Phys.: Condens. Matter **33,** 265701(2021).

[32] Z. Liu, H. Sun, M. Huo, X. Ma, Y. Ji, E. Yi, L. Li, H. Liu, J. Yu, Z. Zhang, Z. Chen, F. Liang, H. Dong, H. Guo, D. Zhong, B. Shen, S. Li, and M. Wang, *Evidence for charge and spin density waves in single crystals of $La_3Ni_2O_7$ and $La_3Ni_2O_6$.* Sci. Chin. Phys. Mech. Astron. **66,** 09AB99 (2022).

[33] Z. Li, W. Guo, T. T. Zhang, J. H. Song, T. Y. Gao, Z. B. Gu, and Y. F. Nie, *Epitaxial growth and electronic structure of Ruddlesden–Popper nickelates ($La_{n+1}Ni_nO_{3n+1}$, n = 1–5).* APL Mater. **8,** 091112 (2020).

[34] H. Li, X. Zhou, T. Nummy, J. Zhang, V. Pardo, W. E. Pickett, J.F. Mitchell, and D.S. Dessau, *Fermiology and electron dynamics of trilayer nickelate $La_4Ni_3O_{10}$.* Nat. Commun. **8,** 704 (2017).

[35] H. Sun, M. Huo, X. Hu, J. Li, Z. Liu, Y. Han, L. Tang, Z. Mao, P. Yang, B. Wang, J. Cheng, D.-X. Yao, G.-M. Zhang, and M. Wang, *Signatures of superconductivity near 80 K in a nickelate under high pressure.* Nature **621,** 493 (2023).

[36] J. Hou, P. T. Yang, Z. Y. Liu, J. Y. Li, P. F. Shan, L. Ma, G. Wang, N. N. Wang, H. Z. Guo, J. P. Sun, Y. Uwatoko, M. Wang, G.-M. Zhang, B. S. Wang, J.-G. Cheng, *Emergence of high-temperature superconducting phase in the pressurized $La_3Ni_2O_7$ crystals*, arXiv:2307.09865 (2023).

[37] Y. Zhang, D. Su, Y. Huang, H. Sun, M. Huo, Z. Shan, K. Ye, Z. Yang, R. Li, M. Smidman, M. Wang, L. Jiao, H. Yuan, *High-temperature superconductivity with*

*zero-resistance and strange metal behavior in La$_3$Ni$_2$O$_7$*, arXiv:2307.14819 (2023).

[38] M. Zhang, C. Pei, Q. Wang, Y. Zhao, C. Li, W. Cao, S. Zhu, J. Wu, Y. Qi, *Effects of Pressure and Doping on Ruddlesden-Popper phases La$_{n+1}$Ni$_n$O$_{3n+1}$*, arXiv:2309.01651(2023).

[39] H. Sakakibara, M. Ochi, H. Nagata, Y. Ueki, H. Sakurai, R. Matsumoto, K. Terashima, K. Hirose, H. Ohta, M. Kato, Y. Takano, K. Kuroki, *Theoretical analysis on the possibility of superconductivity in a trilayer Ruddlesden-Popper nickelate La$_4$Ni$_3$O$_{10}$ under pressure and its experimental examination: comparison with La$_3$Ni$_2$O$_7$,* arXiv:2309.09462 (2023).

[40] G. Wang, N. Wang, J. Hou, L. Ma, L. Shi, Z. Ren, Y. Gu, X. Shen, H. Ma, P. Yang, Z. Liu, H. Guo, J. Sun, G. Zhang, J. Yan, B. Wang, Y. Uwatoko, J. Cheng, *Pressure-induced superconductivity in polycrystalline La$_3$Ni$_2$O$_7$,* arXiv:2309.17378 (2023).

[41] Z. Liu, M. Huo, J. Li, Q. Li, Y. Liu, Y. Dai, X. Zhou, J. Hao, Y. Lu, M. Wang, and W.-H. Wen, *Electronic correlations and energy gap in the bilayer nickelate La$_3$Ni$_2$O$_7$.* arXiv 2307.02950 (2023).

[42] J. Yang, H. Sun, X. Hu, Y. Xie, T. Miao, H. Luo, H. Chen, B. Liang, W. Zhu, G. Qu, C.-Q. Chen, M. Huo, Y. Huang, S. Zhang, F. Zhang, F. Yang, Z. Wang, Q. Peng, H. Mao, G. Liu, Z. Xu, T. Qian, D.-X. Yao, M. Wang, L. Zhao, and X. J. Zhou, *Orbital-Dependent Electron Correlation in Double-Layer Nickelate La$_3$Ni$_2$O$_7$.* arXiv 2309.01148 (2023).

[43] Z. Luo, X. Hu, M. Wang, W. Wú, and D.-X. Yao, *Bilayer Two-Orbital Model of La$_3$Ni$_2$O$_7$ under Pressure.* Phys. Rev. Lett. **131,** 126001 (2023).

[44] Q.-G. Yang, D. Wang, and Q.-H. Wang, *Possible s±-wave superconductivity in La$_3$Ni$_2$O$_7$.* Phys. Rev. Lett. **108,** L140505 (2023).

[45] H. Sakakibara, N. Kitamine, M. Ochi, and K. Kuroki, *Possible high T$_c$ superconductivity in La$_3$Ni$_2$O$_7$ under high pressure through manifestation of a nearly-half-filled bilayer Hubbard model.* arXiv:2306.06039 (2023).

[46] Y. Gu, C. Le, Z. Yang, X. Wu, and J. Hu, *Effective model and pairing tendency in bilayer Ni-based superconductor La$_3$Ni$_2$O$_7$.* arXiv: 2306.07275 (2023).

[47] Y.-F. Yang, G.-M. Zhang, and F.-C. Zhang, *Minimal effective model and possible high-Tc mechanism for superconductivity of La$_3$Ni$_2$O$_7$ under high pressure.* arXiv: 2308.01176 (2023).

[48] F. Lechermann, J. Gondolf, S. Bötzel, and I. M. Eremin, *Electronic correlations and superconducting instability in La$_3$Ni$_2$O$_7$ under high pressure.* arXiv: 2306.05121 (2023).

[49] X. Chen, P. Jiang, J. Li, Z. Zhong, and Y. Lu, *Critical charge and spin instabilities in superconducting La$_3$Ni$_2$O$_7$.* arXiv: 2307.07154 (2023).

[50] Y. Shen, M. Qin, and G.-M. Zhang, *Effective bi-layer model Hamiltonian and density-matrix renormalization group study for the high-Tc superconductivity in La$_3$Ni$_2$O$_7$ under high pressure,* arXiv: 2306.07837 (2023).

[51] B. Geisler, J. J. Hamlin, G. R. Stewart, R. G. Hennig, and P. J. Hirschfeld, *Structural transitions, octahedral rotations, and electronic properties of A$_3$Ni$_2$O$_7$ rare-earth nickelates under high pressure.* arXiv: 2309.15078 (2023).

[52] Y.-B. Liu, J.-W. Mei, F. Ye, W.-Q. Chen, and F. Yang, *The s$^\pm$-Wave Pairing and the*

*Destructive Role of Apical-Oxygen Deficiencies in La$_3$Ni$_2$O$_7$ Under Pressure.* arXiv: 2307.10144 (2023).

[53] S. Kumar, Ø. Fjellvåg, A. O. Sjåstad, and H. Fjellvåg, *Physical properties of Ruddlesden-Popper (n = 3) nickelate: La$_4$Ni$_3$O$_{10}$.* J. Magn. Magn. Mater. **496,** 165915 (2020).

[54] S. Huangfu, X. Zhang, and A. Schilling, *Correlation between the tolerance factor and phase transition in A$_{4-x}$B$_x$Ni$_3$O$_{10}$ (A and B=La, Pr, and Nd; x=0,1,2, and 3).* Phys. Rev. Research **2,** 033247 (2020).

[55] D. F. Segedin, B. H. Goodge, G. A. Pan, Q. Song, H. LaBollita, M.-C. Jung, H. El-sherif, S. Doyle, A. Turkiewicz, N. K. Taylor, J. A. Mason, A. T. N 'Diaye, H. Paik, I. El Baggari, A. S. Botana, L. F. Kourkoutis, C. M. Brooks, and J. A. Mundy, Limits to the strain engineering of layered square-planar nickelate thin films. Nat. Commun. **14,** 1468 (2023).

[56] Z. Zhang and M. Greenblatt, *Synthesis, structure, and properties of Ln$_4$Ni$_3$O$_{10-\delta}$ (Ln =La, Pr, and Nd),* J. Solid State Chem. **117,** 236 (1995).

[57] J. Zhang, D. Phelan, A. S. Botana, Y.-S. Chen, H. Zheng, M. Krogstad, S. G. Wang, Y. Qiu, J. A. Rodriguez-Rivera, R. Osborn, S. Rosenkranz, M. R. Norman, and J. F. Mitchell, *Intertwined density waves in a metallic nickelate.* Nat. Commun. **11,** 6003 (2020).

[58] D. J. Scalapino, *A common thread: The pairing interaction for unconventional superconductors.* Rev. Mod. Phys. **84,** 1383 (2001).

[59] H.-H. Wen, *Developments and Perspectives of Iron-based High-Temperature Superconductors.* Adv. Mater. **20,** 3764–3769 (2008).

[60] M. D. Carvalho, F. M. A. Costa, I. da S. Pereira, A. Wattiaux, J. M. Bassat, J. C. Grenier, and M. Pouchard, *New preparation method of La$_{n+1}$Ni$_n$O$_{3n+1-\delta}$ (n = 2, 3).* J. Mater. Chem. **7,** 2107-2111(1997).

[61] R. W. Cheary, and A. Coelho, *A Fundamental Parameters Approach to X-ray Line Profile Fitting.* J. Appl. Crystallogr. **25,** 109 (1992).

[62] H. K. Mao, J. Xu, and P. M. Bell, *Calibration of the Ruby Pressure Gauge to 800 kbar Under Quasi-Hydrostatic Conditions.* J. Geophys. Res. Solid Earth **91,** 4673-4676 (1986).

[63] C. D. Ling, D. N. Argyriou, G.Wu, and J. J. Neumeier, *Neutron diffraction study of La$_3$Ni$_2$O$_7$: Structural relationships among n = 1, 2, and 3 phases La$_{n+1}$Ni$_n$O$_{3n+1}$,* J. Solid State Chem. **152,** 517 (2000).

[64] M. D. Carvalho, M. M. Cruz, A. Wattiaux, J. M. Bassat, F. M. A. Costa, and M. Godinho, *Influence of oxygen stoichiometry on the electronic properties of La$_4$Ni$_3$O$_{10\pm\delta}$.* J. Appl. Phys. **88,** 544-549 (2000).

[65] J. M. Bassat, C. Allançon, P. Odier, J. P. Loup, M. D. Carvalho, and A. Wattiaux, *Electronic properties of Pr$_4$Ni$_3$O$_{10\pm\delta}$.* Eur. J. Solid State lnorg. Chem. **35,** 173-188 (1998).

[66] Q. Li, J. Si, T. Duan, X. Zhu and H.-H. Wen, *Synthesis, structure, and physical properties of bilayer molybdate Sr$_3$Mo$_2$O$_7$ with flat-band.* Philos. Mag. **100,** 2402-2415 (2020).